\documentclass[%              % paper size definition not needed as JACoW page size is set
               ]{jacow}

\usepackage[english]{babel}

\usepackage{float}

\begin{document}

\title{Using Longitudinal Strong Focusing Principle to Lower Particle Beam Energy Spread Locally in a Storage Ring~\thanks{Work supported by the National Natural Science Foundation of China (No. 12522512),  National
		Key Research and Development Program of China (No. 2022YFA1603401), Beijing
		Outstanding Young Scientist Program (No. JWZQ20240101006) and Tsinghua University Dushi Program. }}

\author[1]{X. J. Deng\thanks{dengxiujie@mail.tsinghua.edu.cn\\ JACoW-IPAC2026-TUP2609}}
\author[1]{A. W. Chao}
\author[1]{W. H. Huang}
\author[1]{Z. L. Pan}
\author[1]{C. X. Tang}
\author[1]{J. Y. Zhao}
%\author[1,2]{A. W. Chao}
%\author[2]{C. Contributor}
\affil[1]{Tsinghua University, Beijing, China }
%\affil[2]{Name of Institute or Affiliation, City, Country}

\maketitle

\begin{abstract}
	In this paper, we propose to use longitudinal strong focusing principle to lower particle beam energy spread locally in a storage ring. An example application of the proposed scheme in reversible Echo SSMB for high-power EUV radiation generation is presented.  We believe strong focusing in the longitudinal dimension has a wide application potential.

\end{abstract}

\section{Introduction}

In some accelerator applications, beam energy spread is the key parameter. For example, in the reversible steady-state microbunching (SSMB) scheme~\cite{Ratner2010SSMB,Ratner2011Reversible,Deng2025Reversible}, if we apply HGHG~\cite{Yu1991} or EEHG~\cite{Stupakov2009} technique for microbunching, a smaller energy spread means a weaker laser energy modulation is needed to obtain the desired bunching at a given harmonic number, or with a given energy modulation strength we can push the bunching to even higher laser harmonics. Correspondingly, it relaxes the technical demand on the laser modulation system, or pushes the SSMB coherent radiation to even shorter wavelength range.   
\begin{figure}[H]
	\centering
	\includegraphics[width=0.825\linewidth]{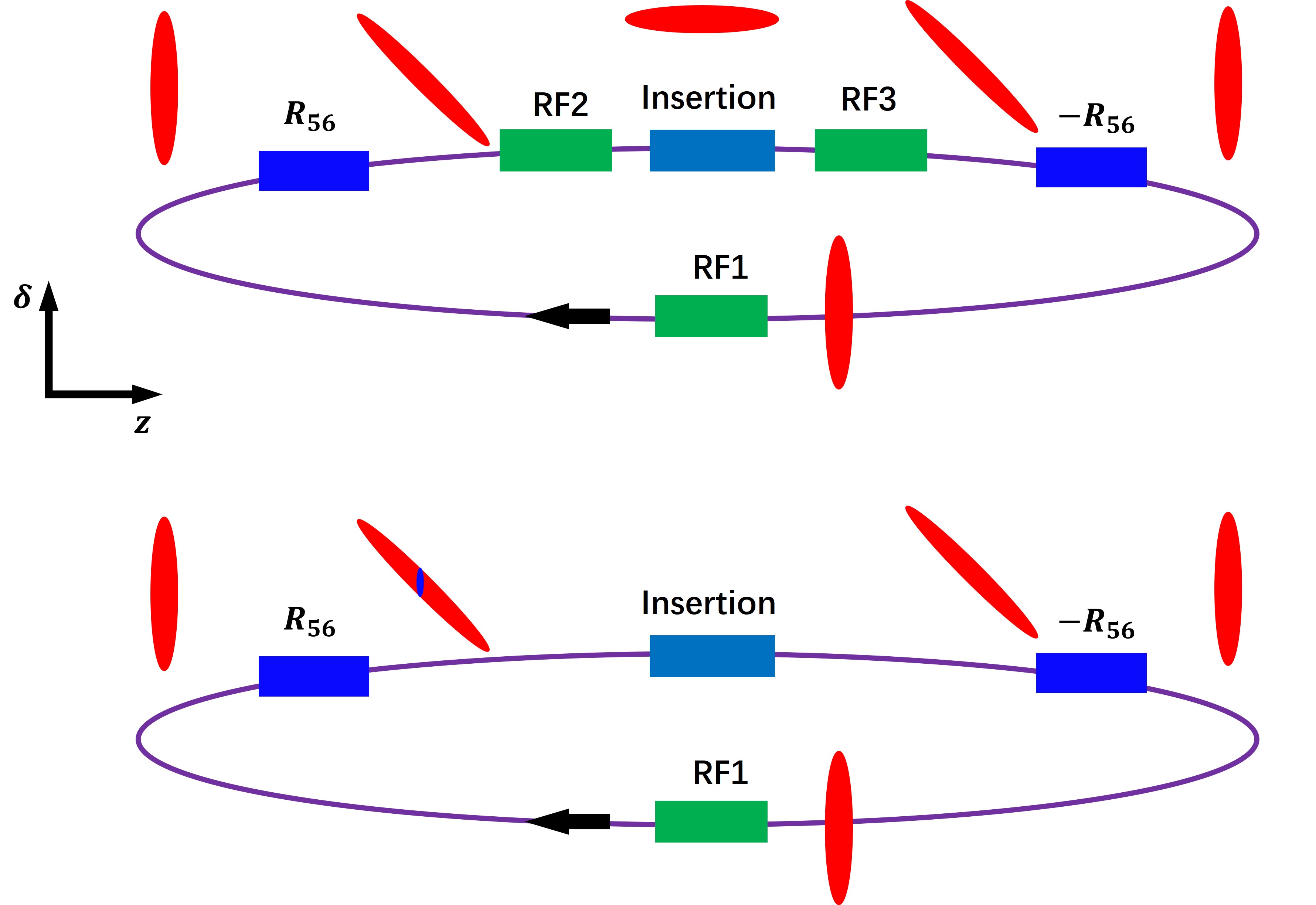}
	\caption{Schematic of applying longitudinal strong focusing principle to lower the global (top) and slice (bottom) beam energy spread at the insertion section in a storage ring.}
	\label{fig:lsfssmblowes}
\end{figure}

From a linear dynamics viewpoint, radio-frequency (RF) cavity functions as a longitudinal quadrupole, and momentum compaction or $R_{56}$ a longitudinal drift. We can use them to manipulate the longitudinal optics like what we usually do in the transverse dimension. Here we propose to use the longitudinal strong focusing (LSF) principle to lower the beam energy spread locally in a storage ring. The schematic of such a proposal is shown in Fig.~\ref{fig:lsfssmblowes}, where we aim at a small energy spread at the insertion section. In the bottom plot only the slice energy spread is lowered, while in the top plot we lower the global energy spread by removing the global energy chirp at the insertion using RF2 and RF3. The proposed scheme applies to both hadron and lepton accelerators, both rings and linacs. In the following, we focus on the case of an electron storage ring. More specifically, we use reversible SSMB for an example application. Other potential applications include storage ring free-electron lasers.

%Here we focus on its application in an electron storage ring, The difference of an electron storage ring is that the equilibrium longitudinal emittance is also given by the ring. 
%
%
% In this paper, we use reversible Echo SSMB as an example for our application. The analysis however can be applied also to other cases.

\section{Longitudinal Strong Focusing}

%\subsection{Electron Storage Ring}

In an electron ring, the equilibrium longitudinal emittance $\epsilon_{z}$ is given by the balance of diffusion and damping,
\begin{equation}
	\frac{d\epsilon_{z}}{dn}=\epsilon_{z}(1-2\alpha_{L})+\Delta\epsilon_{z},
\end{equation}
where $n$ is the revolution number,
$
\alpha_{L}=\frac{J_{s}U_{0}}{2E_{0}}
$
is the longitudinal damping constant, 
with $J_{s}$ being the longitudinal damping partition number which nominally is about 2, $E_{0}$ the particle energy and $U_{0}$ the radiation loss per particle per turn. The growth of $\epsilon_{z}$ from energy diffusion per turn is
\begin{equation}
	\Delta\epsilon_{z}=\oint \frac{d\Delta\sigma_{\delta}^{2}}{ds} (s') \frac{\beta_{z}(s')}{2}ds',
\end{equation}
where $\beta_{z}$ is the longitudinal beta function, whose calculation invokes the longitudinal Courant-Snyder parametrization.

In typical storage ring design, $\beta_{z}$ is basically constant around the ring. Energy excitation at different places contribute to $\epsilon_{z}$  with the same weight, and we have the equilibrium energy spread $
\sigma_{\delta}=\sqrt{\epsilon_{z}/\beta_{z}}=\sqrt{{\Delta\sigma_{\delta}^{2}}/{4\alpha_{L}}}
$, which is independent of $\beta_{z}$, where $\Delta\sigma_{\delta}^{2}$ is the total energy diffusion in one turn.  The situation changes in the proposed scheme, the essence of which is to reduce $\epsilon_{z}$ and $\sigma_{\delta}$ (at given locations) by manipulating $\beta_{z}(s)$, with a given distribution of energy diffusion and radiation damping around the ring.

For a Gaussian bunch with phase space density
$
\psi(z,\delta)=\frac{1}{2\pi\epsilon_{z}}\text{exp}\left(-{J_{z}}/{\epsilon_{z}}\right),
$
where $
J_{z}=\frac{z^{2}+(\alpha_{z}z+\beta_{z}\delta)^{2}}{2\beta_{z}}
$ is the particle longitudinal action, $\epsilon_{z}=\langle J_{z}\rangle$ is the longitudinal beam emittance,
we have the slice energy spread
$
\sigma_{\delta,\text{s}}=\sqrt{{\epsilon_z}/{\beta_z}},
$
the global energy spread
$
\sigma_{\delta,\text{g}}=\sqrt{\epsilon_{z}\gamma_{z}},
$ and the bunch length $\sigma_{z}=\sqrt{\epsilon_{z}\beta_{z}}$. 
%Note that both the above energy spreads are independent of $z$.
%So given a longitudinal emittance, we can lower the slice energy spread by increasing the longitudinal beta function $\beta_{z}$.
%When $\alpha_{z}=0$, the slice energy spread is the same with the global energy spread. 
Note that the energy center of a longitudinal slice is
$
\langle\delta\rangle(z)=-\frac{\alpha_{z}}{\beta_{z}}z.
$
For an appropriate comparison with the case of not applying the LSF scheme, at the insertion we assume $\epsilon_{z,\text{new}}\beta_{z,i,\text{new}}=\epsilon_{z,\text{old}}\beta_{z,i,\text{old}}$ (same bunch length), and $\epsilon_{z,\text{new}}/\beta_{z,i,\text{new}}<\epsilon_{z,\text{old}}/\beta_{z,i,\text{old}}$ (smaller slice energy spread), from which we have $\epsilon_{z,\text{new}}<\epsilon_{z,\text{old}}$ and $\beta_{z,i,\text{new}}>\beta_{z,i,\text{old}}$. To realize a smaller $\epsilon_{z,\text{new}}$, we need to lower $\langle\beta_{z}\rangle$ at the energy diffusion places. This is done by lowering $\beta_{z}$ in the main ring, $\beta_{z,\text{r}}$. So we have a large $\beta_{z,\text{i}}$ in the insertion, but a small $\beta_{z,\text{r}}$ in the main ring. For simplicity, here we have divided the ring into two parts, i.e., the ring part (subscript $r$) and the insertion part (subscript $i$).

%
%Our goal is to realize a small energy spread at the insertion by manipulating $\beta_{z}$ around the ring, with a given distribution of energy diffusion and radiation damping around the ring.  The peak current at the damping wigglers will be correspondingly higher, which is about 81.8 A in our example case shown in the note. There is a lower limit of energy spread reachable by applying this scheme, which is given by Eq. (12).

% The original energy spared is given by
%\begin{equation}
%\sigma_{\delta,\text{old}}=\sqrt{(\Delta\sigma_{\delta,\text{r}}^{2}+\Delta\sigma_{\delta,\text{i}}^{2})/{4\alpha_{L}}},
%\end{equation}
%where we have divided the energy diffusion into two parts, i.e., that of the ring part and that of the insertion part. Now if we apply different $\beta_{z}$ at these two different parts, we have
%\begin{equation}
%\epsilon_{z,\text{new}}=\frac{\Delta\sigma_{\delta,\text{r}}^{2}\beta_{z,\text{r}}/2+\Delta\sigma_{\delta,\text{i}}^{2}\beta_{z,\text{i}}/2}{2\alpha_{L}}
%\end{equation}
%The new slice energy spread at the radiator is 
%\begin{equation}
%\sigma_{\delta,\text{new}}=\sqrt{{\epsilon_{z,\text{new}}}/{\beta_{z,\text{i}}}}
%\end{equation}
%%Note that when $\alpha_{z}=0$ at the insertion, the slice energy spread is also the global energy spread.

We assume the radiation damping constant is not much affected by the LSF. If we want $\sigma_{\delta,i,\text{new}}^{2}=\frac{1}{R}\sigma_{\delta,i,\text{old}}^{2}$ by applying different $\beta_{z}$ at the two parts, 
\begin{equation}
	\begin{aligned}
		{\left(\frac{\Delta\sigma_{\delta,\text{r}}^{2}\beta_{z,\text{r}}/2+\Delta\sigma_{\delta,\text{i}}^{2}\beta_{z,\text{i}}/2}{2\alpha_{L}}\right)}/{\beta_{z,\text{i}}}=\frac{1}{R}\frac{\Delta\sigma_{\delta,\text{r}}^{2}+\Delta\sigma_{\delta,\text{i}}^{2}}{4\alpha_{L}},
	\end{aligned}
\end{equation}
where $\Delta\sigma_{\delta,\text{r,i}}^{2}$ are the energy diffusion in the two parts, then
\begin{equation}
	\begin{aligned}
		\frac{\beta_{z,\text{r}}}{\beta_{z,\text{i}}}=\frac{1}{R}+\left(\frac{1}{R}-1\right)\frac{\Delta\sigma_{\delta,\text{i}}^{2}}{\Delta\sigma_{\delta,\text{r}}^{2}}.
	\end{aligned}
\end{equation}

To make $\sigma_{\delta,\text{new}}<\sigma_{\delta,\text{old}}$, which means $R>1$, we need $\beta_{z,\text{i}}>\beta_{z,\text{r}}$. 
The theoretical upper limit of $R$ is 
\begin{equation}
	R_{\text{max}}=1+\frac{\Delta\sigma_{\delta,\text{r}}^{2}}{\Delta\sigma_{\delta,\text{i}}^{2}}.
\end{equation}

The scheme can also be used to enlarge the energy spread locally in a ring, and there is no upper limit of such widening. Apart from the energy spread, we remind that the LSF principle can also be used to manipulate the bunch length. More details about the longitudinal Courant-Snyder formalism and emittance minimization can be found in Refs.~\cite{Deng2021Courant,Zhang2021Ultra,Deng2026NST}. 

%We assume that If we want to remove the global energy chirp at the insertion, which means $\alpha_{z}=0$ also at the insertion, we need
%\begin{equation}
%\begin{aligned}
%h_{\text{RF2}}&=-\frac{R_{56}}{\beta_{z,\text{r}}^2+R_{56}^{2}}=-\frac{R_{56}}{\beta_{z,\text{i}}\beta_{z,\text{r}}}.
%\end{aligned}
%\end{equation}

\section{A LSF Echo SSMB EUV Source}

\begin{figure}[tb]
	\centering
	\includegraphics[width=0.85\linewidth]{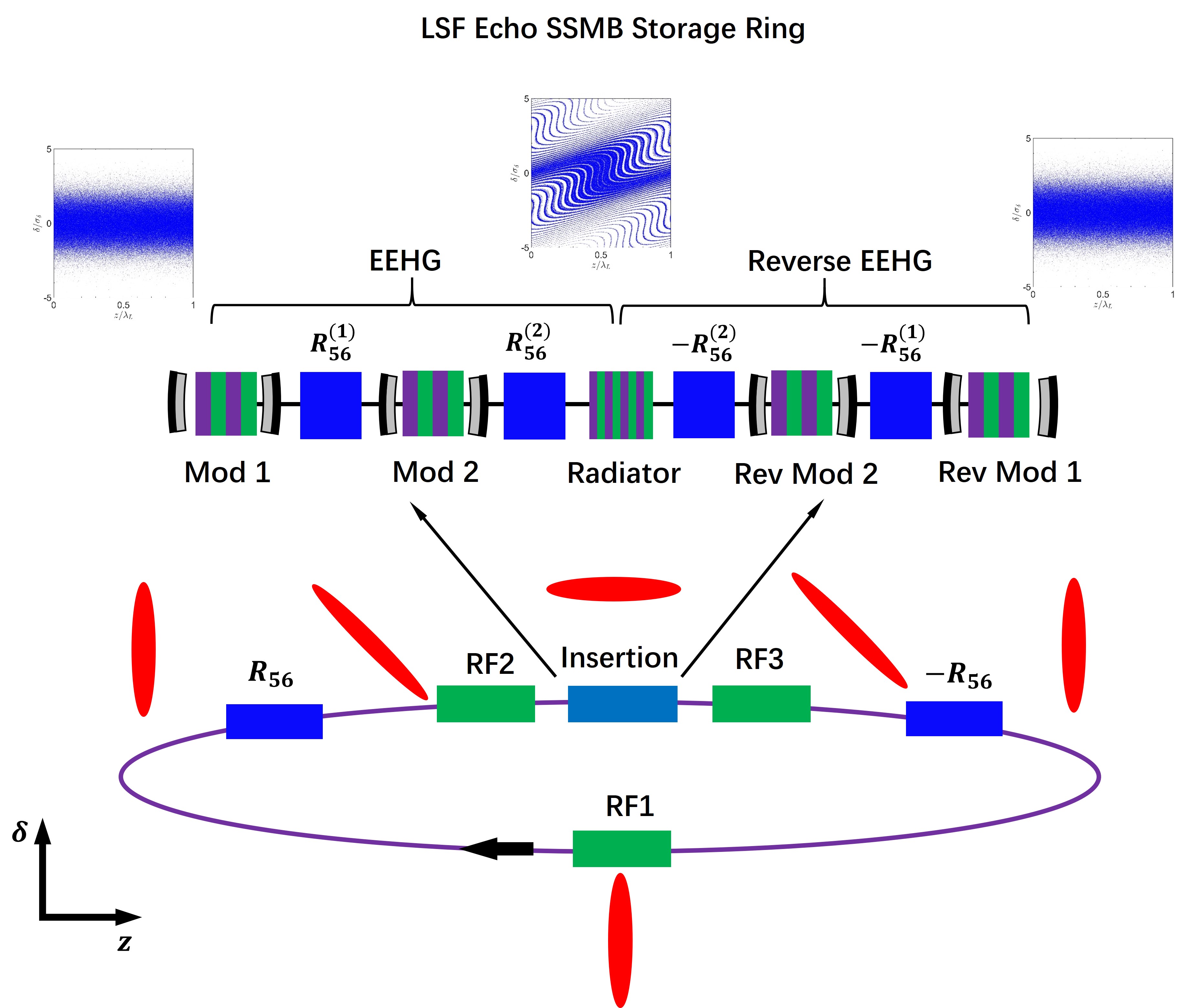}
	\caption{A schematic layout of a LSF Echo SSMB ring.}
	\label{fig:lsfechossmb}
\end{figure}

Now we apply the proposed scheme in reversible SSMB. To obtain high-harmonic bunching for short-wavelength such as EUV coherent radiation, we invoke EEHG as the microbunching scheme. We call such a combination of novel ideas as LSF Echo SSMB. The schematic layout and some key parameters of such a ring for high-power EUV radiation are shown in Fig.~\ref{fig:lsfechossmb} and Table~\ref{tab:LSFEchoSSMB}. 

% A small energy spread is desired in such a light source as mentioned in the introduction.

Note that in reversible SSMB, the electron beam is only microbunched in the insertion section, and outside the insertion it is just like a conventional RF bunch. In such a ring, the diffusion of energy are mainly from: quantum excitation of dipoles and damping wigglers, and the energy spread growth due to non-perfect laser modulation cancellation of the insertion, which is the EEHG, radiator, plus the reverse EEHG section as shown in Fig.~\ref{fig:lsfechossmb}. The radiation damping has two contributions: dipoles and damping wigglers. In our example, for both diffusion and damping, the dipole contribution is negligible compared to that of the damping wigglers, and we have $\Delta\sigma_{\delta,\text{r}}^{2}\approx\Delta\sigma_{\delta,\text{w}}^{2}$ and $U_{0}\approx U_{0\text{w}}$.

\begin{table}[tb]
	\caption{\label{tab:LSFEchoSSMB}
		Parameters of a LSF Echo SSMB EUV source.}
	\centering
	%\begin{ruledtabular}
	\begin{tabular}{lll}  
		\hline
		Para. & \multicolumn{1}{l}{\textrm{Value}}  & Description \\
		%\colrule
		\hline 
		%		{\bf Basic}\\
		%\hline
		$E_{0}$ & 600 MeV & Beam energy \\	
		$I_{A} $ & 1 A & Average current \\
		%		&$E_{\text{low}}$ & ${\color{red}299\ \text{MeV}}$  & Beam energy \\	
		$|\eta C_{0}|$ & 0.098 m & Ring momentum compaction \\
		$|R_{56}|$ & 3.04 m & $\pm R_{56}$ before/after insertion \\
		RF1 & 1.2 MV & 500 MHz RF in main ring \\
		RF2,3 & 4.07 MV & 1.5 GHz RF for insertion \\
		%&$I_{P}$ & $8$ A & Peak current \\
		%&$f$ & 12.5\% & Beam filling factor \\		
		%$I_{P,\text{i}}$ & 50 A & Peak current in the insertion\\
		%&$I_{P,\text{r}} (I_{A})$ & 92.45 (1) A & Peak (average) current \\
		%		&$I_{A}$ & $1$ A & Average current \\
		%$\eta$ & $1\times10^{-6}$ & Phase slippage \\
		%&$B_{\text{r}}$ & 1.33 T & Bending field in the ring \\
		%&$\rho_{\text{r}}$ & 1 m & Bending radius in the ring \\
		%&$U_{0\text{r}}$ &  7.76 keV & Radiation loss per particle from ring dipoles\\
		%$\theta$ & $\frac{\pi}{12}$ & Bending angle of each dipole in the ring \\	
		%$\sigma_{z,\text{min}}$ & 42 nm & Theoretical minimum bunch length\\
		%$\epsilon_{z,\text{min}}$ & 41 pm & Theoretical minimum longitudinal emittance\\
		%$h_{\text{r}}$ & 500 m$^{-1}$ & Energy chirp strength in the ring\\
		%$\epsilon_{x}$ & 2.7 nm & Natural horizontal emittance \\	
		%$\tau_\delta$ & 14.7 ms & Longitudinal damping time\\	
		%		&$\sigma_{\delta{\text{0}}}$ &  $4.2\times10^{-4}$ & Natural energy spread\\
		%		&$\sigma_{\delta{\text{0,\text{w}}}}$ &  $4.9\times10^{-4}$ & Natural energy spread with damping wiggler\\
		$B_{0{\text{w}}}$ &  {4 T} & Field strength of wigglers\\
		$L_{{\text{w}}}$ &  {100 m} & Total length of wigglers\\
		$U_{0\text{w}}$ &  {375 keV} & Radiation loss per particle\\
		%&$\sigma_{\delta0}$ &  $4.2\times10^{-4}$ & Natural energy spread with only dipoles\\
		%&$\sigma_{\delta,\text{Wiggler}}$ &  $7.2\times10^{-4}$ & Energy spread with damping wiggler\\
		%&$\sigma_{\Delta z}$ & {\color{red} 0.5 nm} & Effective smearing of $z$\\
		$\Delta\sigma_{\delta,\text{w}}^{2}$ & $1.1\times10^{-9}$ & Growth of $\sigma_{\delta}^{2}$ in wigglers\\
		%&$\Delta\sigma_{\delta,\text{IBS}}^{2}$ & $2\times10^{-9}$ & Single-pass growth of energy spread from other\\
		%&$\Delta\sigma_{\delta,\text{other}}^{2}$ & $2.6\times10^{-9}$ & Single-pass growth of energy spread squared\\
		$\Delta\sigma_{\delta,\text{i}}^{2}$ & $1\times10^{-9}$ & Growth of $\sigma_{\delta}^{2}$ in insertion\\
		%$\sigma_{\delta,i,\text{old}}$ &  $9.25\times10^{-4}$ & Energy spread without LSF\\
		%$R$ & $1.52$ & Reduction factor of $\sigma_{\delta}^{2}$\\
		%&$J_{s}$ & $2$ & Longitudinal damping partition number\\
		$\sigma_{\delta,i,\text{new}}$ &  $7.5\times10^{-4}$ & Energy spread at insertion\\
		%& $\Delta\delta_{CUR}$ & $\Delta\delta_{CUR}=0.2\Delta\delta_{mean}\xi$ & Impact of coherent undulator radiation\\
		%&$\sigma_{\Delta z,\text{Partial Alpha}}$ & 0.13 nm & Smearing of $z$ from radiator quantum excitation\\	
		\hline 
		%		{\bf For EUV (13.5 nm)}\\
		%\hline 
		$\lambda_{L}$ & 1030 nm & Modulation laser wavelength	\\
		$A_{1}$ & $5\times10^{-4}$  & First-stage modulation \\ 
		%&$h_{1}$ &  $10880\ \text{m}^{-1}$ & First-stage energy chirp strength \\ 
		%&$R_{56}^{(1)}$ &  $6.3\ \text{mm}$ & First-stage $R_{56}$ \\ 
		%&$\lambda_{u\text{Mod1}}$ &  7.69 cm &  Modulator undulator period\\  
		%		&$B_{0\text{Mod1}}$ &  1.2 T &  Modulator peak magnetic flux density\\ 
		%		%&$K_{u\text{Mod1}}$ &  6.44 & $K$ of modulator undulator\\ 
		%		%&$N_{u\text{Mod1}}$ &  40 & $N_{u}$ of modulator undulator \\ 
		%&$L_{u\text{Mod1}}$ &  0.989 m (\textcolor{red}{$N_{u}=13$}) & Modulator length (two sub modulators)	\\ 
		%&$h_{1}R_{56\text{Mod1}}$ &  ${\color{red}0.1}$ & Laser modulator one \\ 
		%		&$Z_{R}=\frac{L_{u}}{3}$ &  0.5 m & Rayleigh length \\ 
		%& $w_{0}(Z_{R}=\frac{L_{u}}{3})$ & 582 $\mu$m & Laser beam waist\\
		%&$P_{L1}(Z_{R}=\frac{L_{u}}{3})$ &  4.8 MW & Modulation peak laser power\\ 
		%&$f_{1}$ & 12.5\% & Beam filling factor \\
		$P_{1}$ &  ${\sim100\ \text{kW}}$ & Average laser power\\ 
		%& $PD_{mirror1}$ & 5.25 MW/cm$^2$ & Power density onmirrors  (7.5 m distance)\\
		%\hline 
		%		{\bf For EUV (13.5 nm)}\\
		%\hline 
		%&$\lambda_{2}$ & 1030 nm & Modulation laser wavelength	\\
		$A_{2}$ &  $1.5\times10^{-3}$ & Second-stage modulation \\ 
		%&$h_{2}$ &  $11813\ \text{m}^{-1}$ & Second-stage energy chirp strength \\ 
		%&$R_{56}^{(2)}$ &  $85\ \mu\text{m}$ & Second-stage $R_{56}$ \\ 
		%&$\lambda_{u\text{Mod2}}$ &  6.07 cm &  Modulator undulator period\\  
		%		&$B_{0\text{Mod2}}$ &  1.2 T &  Modulator peak magnetic flux density\\ 
		%		%&$K_{u\text{Mod1}}$ &  6.44 & $K$ of modulator undulator\\ 
		%		%&$N_{u\text{Mod2}}$ &  40 & $N_{u}$ of modulator undulator \\ 
		%&$L_{u\text{Mod2}}$ &  1.826 m /2 (\textcolor{red}{$N_{u}=24/2$}) & Modulator length	\\ 
		%		%&$h_{2}R_{56\text{Mod2}}$ &  ${\color{red}0.0074}$ & Laser modulator two \\ 
		%		%		&$Z_{R}=\frac{L_{u}}{3}$ &  0.5 m & Rayleigh length \\ 
		%		%& $w_{0}(Z_{R}=\frac{L_{u}}{3})$ & 582 $\mu$m & Laser beam waist\\
		%		%&$P_{L2}(Z_{R}=\frac{L_{u}}{3})$ &  258 kW & Modulation peak laser power\\ 
		%		%&$f_{1}$ & 12.5\% & Beam filling factor \\
		$P_{2}$ &  ${\sim1\ \text{MW}}$ & Average laser power~\cite{Lu2024OEC}\\ 
		%		& $PD_{mirror2}$ & 2 MW/cm$^2$ & Power density on mirrors  (7.5 m distance)\\
		\hline
		%		{\bf For EUV (13.5 nm)}\\
		%\hline 
		%&$\lambda_{3}$ & 343 nm & Modulation laser wavelength	\\
		%&$V_{3}$ &  \textcolor{red}{$0.06V_{2}$} & Third harmonic upstream \\ 
		%	&$P_{3}$ &  \textcolor{red}{6 kW} & Second harmonic downstream \\
		%&$h_{2}$ &  $11813\ \text{m}^{-1}$ & Second-stage energy chirp strength \\ 
		%&$P_{LA2}(Z_{R}=\frac{L_{u}}{3})$ &  ${\color{red}2.5\ \text{kW}}$ & Average modulation laser power\\ 
		%		& $PD_{mirror2}$ & 2 MW/cm$^2$ & Power density on mirrors  (7.5 m distance)\\
		%		\hline
		%		%		{\bf For EUV (13.5 nm)}\\
		%		%\hline 
		%		&$\lambda_{2,3}$ & 343 nm & Modulation laser wavelength	\\
		%		%&$A_{2}$ &  \textcolor{red}{13.6 kV} & Second-stage energy modulation strength \\ 
		%		%&$h_{2}$ &  $11813\ \text{m}^{-1}$ & Second-stage energy chirp strength \\ 
		%		&$P_{LA2}(Z_{R}=\frac{L_{u}}{3})$ &  ${\color{red}3\ \text{kW}}$ & Average modulation laser power\\ 
		%		%		& $PD_{mirror2}$ & 2 MW/cm$^2$ & Power density on mirrors  (7.5 m distance)\\
		%	\hline
		$\lambda_{R}$ & $13.55$ nm & Radiation wavelength\\
		$\epsilon_{\bot}$ & 1 nm & Transverse emittance \\ 	
		$b_{76}$ & 0.058 & Bunching factor\\
		%&$\sigma_{\bot}$ & 10 $\mu$m & Transverse electron  beam size\\
		$\lambda_{u\text{Rad}}$ &  2 cm & Radiator undulator period\\
		%$B_{0\text{Rad}}$ &  0.7 T &  Radiator field strength\\   
		%		&$K_{u\text{Rad}}$ &  1.12 & $K$ of radiator undulator\\ 
		%		&$N_{u\text{Rad}}$ &  $238$ & Number of undulator periods\\ 
		$L_{u\text{Rad}}$ &  2 m  & Radiator length	\\
		%& $\beta_{\bot,\text{center}}$ & $0.4L_{u\text{Rad}}$ & $\beta_{\bot}$ in radiator center\\
		%		&$\sigma_{\bot}(\text{Rad})=\sqrt{\f6rac{\epsilon_{\bot}L_{u}}{3}}$ & 77 $\mu$m & Effective transverse beam size\\ 
		%&$P_{P}$ & 8 kW & Peak radiation power \\
		%&$P_{A}$ & ${\color{red}100\ \text{W}} $ & Average radiation power {\color{red} without third harmonic laser}\\
		$P_{A\text{Rad}}$ & $1\ \text{kW} $ & Average radiation power \\
		\hline 
		%		&$L_{G,1D}$ & ${\color{red}1.01\ \text{m}} $ & 1D FEL gain length \\
		%		&$L_{G,3D}$ & ${\color{red}15.2\ \text{m}} $ & 3D FEL gain length \\
		%		\hline 
		%		&$\lambda_{R}=\frac{\lambda_{L}}{152}$ & $6.78$ nm & Radiation wavelength\\
		%		&$\epsilon_{\bot}$ & 0.7 nm & Effective transverse emittance \\ 	
		%		&$b_{76}$ & 0.014 & Bunching factor\\
		%		%&$\sigma_{\bot}$ & 10 $\mu$m & Transverse electron  beam size\\
		%		&$\lambda_{u\text{Rad}}$ &  1.68 cm & Radiator undulator period\\
		%		&$B_{0\text{Rad}}$ &  0.3 T &  Radiator peak magnetic flux density\\   
		%		%		&$K_{u\text{Rad}}$ &  1.12 & $K$ of radiator undulator\\ 
		%		%		&$N_{u\text{Rad}}$ &  $238$ & Number of undulator periods\\ 
		%		&$L_{u\text{Rad}}$ &  2.95 m ($N_{u}=176$)  & Radiator length	\\
		%		%		&$\sigma_{\bot}(\text{Rad})=\sqrt{\frac{\epsilon_{\bot}L_{u}}{3}}$ & 77 $\mu$m & Effective transverse beam size\\ 
		%		%&$P_{P}$ & 8 kW & Peak radiation power \\
		%		%&$P_{A}$ & ${\color{red}100\ \text{W}} $ & Average radiation power {\color{red} without third harmonic laser}\\
		%		&$P_{A}$ & ${\color{red}23\ \text{W}} $ & Average radiation power\\
		%		\hline 
		%		& $R_{1}$ & 0.645 & Compensation ratio of first undulator $R_{56}$ \\
		%		& $R_{2}$ & 0.653 & Compensation ratio of second undulator $R_{56}$ \\
		%		& $R_{3}$ & 0.653 & Compensation ratio of third undulator $R_{56}$ \\
		%		& $R_{4}$ & 0.645 & Compensation ratio of fourth undulator $R_{56}$ \\
		%		\hline
	\end{tabular}
	%\end{ruledtabular}
\end{table}

For quantum excitation of the damping wigglers, we have 
\begin{equation}
	\begin{aligned}
		\Delta\sigma_{\delta,\text{w}}^{2}&=\frac{55\alpha\hbar^{2}c^{2}\gamma^{7}}{24\sqrt{3}E_{0}^{2}}\int\frac{ds'}{|\rho(s')|^{3}}
		%&\approx\frac{55\alpha\hbar^{2}c^{2}\gamma^{7}}{24\sqrt{3}}\frac{1}{E_{0}^{2}} \frac{4}{3\pi} L_{u} \frac{1}{\rho_{0}^{3}}\\
		=1.1\times10^{-9},
	\end{aligned}
\end{equation}
with $\alpha$ the fine structure constant, $c$ the speed of light in free space, $\hbar$ the reduced Planck constant, $\gamma$ the Lorentz factor, $\rho$ the bending radius.
The single-pass energy spread growth from the insertion is assumed to be
$
\Delta\sigma_{\delta,\text{i}}^{2}=1\times10^{-9}.
$ We recognize it requires dedicated lattice design efforts~\cite{Pan2025Isochronous,Zhao2026IPAC} and intrabeam scattering (IBS) optimization~\cite{Tang2026Review} to realize such a mild level of energy spread growth from the insertion. Without applying the LSF scheme, the equilibrium energy spread is
$
\sigma_{\delta,\text{old}}=9.25\times10^{-4}.
$ 
%The bunching factor in EEHG is basically determined by $A_{1}/\sigma_{\delta}$ and the harmonic number, with $A_{1}=eV_{L1}/E_{0}$ the first-stage energy modulation strength.
Now we use the proposed LSF scheme to lower the energy spread at the insertion. With the given values of $\Delta\sigma_{\delta,\text{i}}^{2}$ and $\Delta\sigma_{\delta,\text{r}}^{2}$, we have 
$
R_{\text{max}}=2.1.
$ To avoid too dramatic requirements on $\beta_{z}$, $R_{56}$ and RF cavities, we may choose $R=1.52$, then the new equilibrium energy spread at the insertion is 
$
\sigma_{\delta,i,\text{new}}={\sigma_{\delta,\text{old}}}/{\sqrt{R}}=7.5\times10^{-4}.
$
Correspondingly, we need
${\beta_{z,\text{i}}}/{\beta_{z,\text{r}}}=2.88$.

%${\beta_{z,\text{r}}}/{\beta_{z,\text{i}}}=0.347,$ or

%\subsection{Required $R_{56}$, $\eta C_{0}$}
Assume the RF cavity applied in the main ring (RF1) has a frequency of 500 MHz.
%Assuming that we have a Gaussian bunch with an RMS length of $\sigma_{t,\text{i}}$ and peak current of $I_{P,\text{i}}$, then the bunch charge is given by
%$Q=I_{P,\text{i}}\sqrt{2\pi}\sigma_{t,\text{i}}.$
Further assume a peak current of 50~A at the insertion, and 100\% filling factor of RF buckets,  we need a bunch length $\sigma_{t,\text{i}}=16$ ps at the insertion to get an average beam current of 1 A.
The longitudinal beta function at the insertion is then
$
\beta_{z,\text{i}}=\frac{c\sigma_{t,\text{i}}}{\sigma_{\delta,i,\text{new}}}=6.4\ \text{m}.
$
Correspondingly
$
\beta_{z,\text{r}}=2.22\ \text{m}.
$
We assume that $\alpha_{z,\text{r}}\approx0$ which can be realized in practice, then 
$
\beta_{z,\text{i}}=\beta_{z,\text{r}}+\frac{R_{56}^2}{\beta_{z,\text{r}}}.
$
So we have
$
|R_{56}|=\beta_{z,\text{r}}\sqrt{{\beta_{z,\text{i}}}/{\beta_{z,\text{r}}}-1}=3.04\ \text{m},
$
which is a large but realizable value.
If RF1 has a voltage of 1.2 MV, then we have the energy chirp strength around the synchronous phase
$
h_{\text{RF1}}=\frac{eV_{\text{RF1}}}{E_{0}}k_{\text{RF1}}\cos\phi_{s}=1.99\times10^{-2}\ \text{m}^{-1},
%=\frac{1.2}{600}\frac{2\pi}{\frac{3\times10^{8}}{0.5\times10^{9}}}\sqrt{1-\left(\frac{0.375}{1.2}\right)^{2}}\text{m}^{-1}
$ where the synchronous phase is defined according to $eV_{\text{RF1}}\sin\phi_{s}=U_{0}$.
%\begin{equation}
%\sin\phi_{s}=\frac{0.375}{1.2}\Rightarrow \phi_{s}=\pi-\arcsin\left(\frac{0.375}{1.2}\right)=2.8238\ \text{rad}
%\end{equation}
We assume the global synchrotron tune of the ring is still much less than 1, then
$
\beta_{z,\text{r}}\approx\sqrt{{\eta C_{0}}/{h_{\text{RF1}}}}
$ with $\eta$ the phase slippage and $C_{0}$ the circumference of the ring~\cite{DengSpringer2024},
and we have
$
\eta C_{0}=h_{\text{RF1}}\beta_{z,\text{r}}^{2}=98.1\ \text{mm},
$
which should also be straightforward to realize. 
The RF bucket half-height is $\hat{\delta}_{\frac{1}{2}}=\frac{2}{k_{\text{RF1}}\beta_{z,\text{r}}}|1-\left(\frac{\pi}{2}-\phi_{s}\right)\tan\phi_{s}|=5.06\times10^{-2}$.

% which should be large enough. 

%The bucket half height is $5\times10^{-2}$.

%\begin{equation}
%\begin{aligned}
%\hat{\delta}_{\frac{1}{2}}=5.06\times10^{-2}
%%&=\frac{2\nu_{s}}{h_{\text{har}}|\eta|}\bigg|1-\left(\frac{\pi}{2}-\phi_{s}\right)\tan\phi_{s}\bigg|\\
%&=\frac{2}{k_{RF}\beta_{z,\text{r}}}\bigg|1-\left(\frac{\pi}{2}-\phi_{s}\right)\tan\phi_{s}\bigg|,
%\end{aligned}
%\end{equation}

Note that the peak current in the ring section is given by
$
I_{P,\text{r}}=\sqrt{{\beta_{z,\text{i}}}/{\beta_{z,\text{r}}}}I_{P,\text{i}}=84.85\ \text{A}.
$
We recognize more work on collective effects is needed to justify the applied beam current. There are many novel features concerning such studies in the proposed scheme and in SSMB in general, such as the breakdown of adiabatic condition due to the significant phase space evolution around the ring, the exchange of the beam head and tail when traveling around the ring, the impact of coherent radiation and turn-by-turn laser modulation on collective instabilities. Readers can refer to Refs.~\cite{Bian2025Threshold,Bian2026IPAC,Pan2025IBS,Zhao2025Method,Tsai2025Simple,Tsai2022Theoretical,Dai2026Longitudinal} for more details about the ongoing efforts in this respect.

%\subsection{Requirement on RF Cavities}
%
%Now we check the requirement on the lattice, and for case in Fig.~\ref{fig:lsfssmblowes} also the RF cavities for energy chirping and dechirping.

\iffalse

Using $\left(\begin{matrix}
	z\\
	\delta
\end{matrix}\right)$ as the phase space coordinate, the transfer matrix from ring to insertion is

\begin{equation}
	\begin{aligned}
		{\bf M}	=\left(
		\begin{matrix}
			1&0\\
			h&1
		\end{matrix}
		\right)\left(
		\begin{matrix}
			1&R_{56}\\
			0&1
		\end{matrix}
		\right)=\left(
		\begin{matrix}
			1&R_{56}\\
			h&1+hR_{56}
		\end{matrix}
		\right)
	\end{aligned}
\end{equation}
The initial Twiss matrix is given by
\begin{equation}
	{\bf T}_{i}=\left(
	\begin{matrix}
		\beta_{z}&0\\
		0&\frac{1}{\beta_{z}}
	\end{matrix}
	\right)
\end{equation}
Then the final Twiss matrix is given by 
\begin{equation}
	{\bf T}_{f}={\bf M}{\bf T}_{f}{\bf M}^{T}=
	\begin{pmatrix}
		\beta_z + \dfrac{R_{56}^2}{\beta_z} &
		h\beta_z + \dfrac{R_{56}}{\beta_z} + \dfrac{hR_{56}^2}{\beta_z} \\[2ex]
		h\beta_z + \dfrac{R_{56}}{\beta_z} + \dfrac{hR_{56}^2}{\beta_z} &
		h^2\beta_z + \dfrac{(1 + hR_{56})^2}{\beta_z}
	\end{pmatrix}
\end{equation}
We hope that the final $\alpha_{z}=0$, then we have
\begin{equation}
	h\beta_z + \dfrac{R_{56}}{\beta_z} + \dfrac{hR_{56}^2}{\beta_z}=0\Rightarrow h=-\frac{R_{56}}{\beta_{z}^2+R_{56}^{2}}=-\frac{R_{56}}{\beta_{zi}\beta_{zf}}.
\end{equation}
\fi
Now if we want to further make $\alpha_{z,\text{i}}=0$, we can use another RF cavity (RF2) to remove the global energy chirp. To make the dynamics repeats turn by turn, we can apply a negative $R_{56}$ and reverse RF kick (RF3) following the insertion section. In principle, we can also use in the downstream the same sign $R_{56}$ and RF kick with that of the upstream, although the nonlinear longitudinal nonlinear dynamics is more subtle then. The required linear chirp strength for RF2 and RF3 is
$
h_{\text{RF2,3}}=|{R_{56}}/{\beta_{z,\text{i}}\beta_{z,\text{r}}}|=0.214\ \text{m}^{-1}.
$
Now if we use 1.5~GHz cavities for RF2 and RF3, the required cavity voltage is
$
V_{\text{RF2,3}}={E_{0}h_{\text{RF2,3}}}/{ek_{\text{RF2,3}}}=4.07\ \text{MV},
$
which should be doable using superconducting technology. 

%Actually, it we can realize higher RF voltage at lower frequency, it also works for our purpose. Here we just use X-band RF cavity as an example. 

%\begin{equation}
%\Delta\sigma_{\delta,\text{i}}^{2}\propto V_{1}^{2}
%\end{equation}
%
%And the bunching factor will be proportional to $A_{1}/\sigma_{\delta}$
%
%
%
%Helical undulator as radiator. 
%
%Gaussian bunch should be considered!

The damping wigglers are mainly used to speed up damping and increase the tolerance of the non-perfect modulation cancellation. To control the quantum excitation of wigglers to horizontal emittance~\cite{Deng2026NST}, we need to optimize the $\mathcal{H}_{x}$ inside the wigglers and the wiggler period length need to be
$
\lambda_{w}[\text{m}]\leq 3.19\sqrt{{N_{wc}J_{x}E_{0}[\text{GeV}]\epsilon_{x}[\text{nm}]}/{B_{0w}^{3}[\text{T}]L_{w}[\text{m}]}},
$
where we have assumed there are $N_{wc}$ identical wigglers, with a total length of $L_{w}$.
%\frac{N_{c}^{\frac{1}{2}}J_{x}^{\frac{1}{2}}E^{\frac{1}{2}}_{0}[\text{GeV}]\epsilon^{\frac{1}{2}}_{x0}[\text{nm}]}{B_{0w}^{\frac{3}{2}}[\text{T}]L_{w}^{\frac{1}{2}}[\text{m}]}
Nominally the horizontal damping partition number $J_{x}\approx1$. For our example, $\epsilon_{x}=1$ nm,  $B_{0w}=4$ T, $L_{w}=100$~m, ${N_{wc}=40}$ which means each small wiggler has a length of 2.5~m, we need
$
\lambda_{w}\leq0.195\ \text{m}.
$
A wiggler with a period of about 15 cm and peak field of 4~T is feasible using superconducting magnet technology~\cite{Shkaruba2018Status}.

\section{Methods to Enhance Radiation}

To obtain high radiation power, we need to optimize the beam 6D phase space distribution at the radiator which in our case is  an undulator. Under far-field and small opening angle approximation, the total field of radiation close to the undulator resonance frequency is given by
%\vspace{-8pt}
	\begin{equation}
		%\begin{small}
		\begin{aligned}			
			&E_{\text{total}}(\boldsymbol{\theta},\omega)\approx N_{e}\int  E_{\text{ref,point}}({\bf 0},\omega)e^{-i\frac{\omega}{c}(\theta_{x}x+\theta_{y}y+z)}\\
			&\text{sinc}\left[\frac{\omega L_{u}[(\theta_{x}-x')^{2}+(\theta_{y}-y')^2]}{4c}-2\pi N_{u}\delta\right] \psi({\bf X}) d{\bf X},		
		\end{aligned}
	%\end{small}
	\end{equation}
in which ${\bf X}=(x,x',y,y',z,\delta)^{T}$, $\psi({\bf X})$ is the normalized beam distribution at the radiator center where we assume the maximal bunching factor is reached, $L_{u}$ and $N_{u}$ are the undulator length and period number, respectively.  The radiation spatial-spectral distribution is then $\frac{d^{2}W}{d\omega d\Omega}=\frac{\epsilon_{0}c}{\pi}|RE_{\text{total}}(\boldsymbol{\theta},\omega)|^{2}$, with $\epsilon_{0}$ vacuum permittivity, $(\theta_{x},\theta_{y})=(\theta\cos\varphi,\theta\sin\varphi)$, $d\Omega=\theta d\theta d\varphi$. With this formulation, we find $\beta_{x,y}$ at the radiator center should match the undulator and the optimal value is about $0.4L_{u}$. Considering the impact of energy spread and laser-induced energy modulations, there is an optimal $N_{u}\approx1/5\sigma_{\delta,\text{effective}}$, above which the coherent radiation power will drop instead of growing. 

Apart from the phase space matching, we can detune the undulator to make the on-axis resonance wavelength slightly  blue-shifted with respect to our desired 13.5~nm radiation. This will result in an increase of the overall 13.5 nm radiation power due to a larger contribution of the solid angle $\theta d\theta$ from the off-axis red-shifted radiation. The power enhancement can be close to a factor of 2, if the transverse dimension of the electron beam is ignored. But the transverse beam emittance will make this benefit less pronounced. Further, we can add a weak high-harmonic laser modulation at the second-stage modulation of EEHG to boost the bunching factor (a third-harmonic laser is assumed in Table~\ref{tab:LSFEchoSSMB}).

\section{Fold the Ring Arc Section}

\begin{figure}[H]
	\centering
	\includegraphics[width=0.8\linewidth]{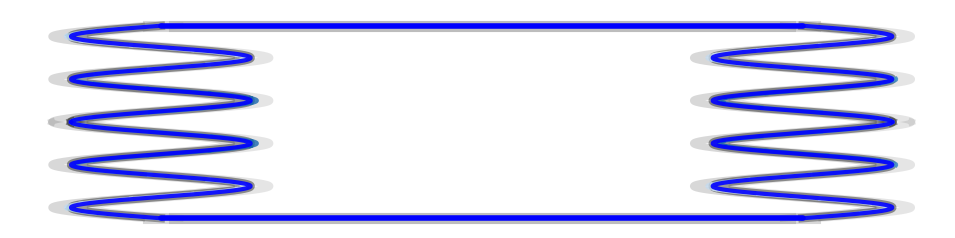}
	\caption{Fold the ring arc sections for a compact footprint.  }
	\label{fig:HelicalArcForDamping}
\end{figure}

One notable feature in our example parameters is the strong and long damping wigglers. The $40\times2.5$~m wigglers are mainly placed in the ring arc sections. Considering the transverse optics control, nonlinear dynamics and IBS optimization, the arc section length is $>200$ m. To save the ring footprint, we can fold the arc section into helical arcs as shown in Fig.~\ref{fig:HelicalArcForDamping}. One straight is for the SSMB insertion, and the other for energy compensation, injection and extraction, etc. There could be many variants of such a folding layout.
%\vspace{-5pt}

%The
%two straight sections are at different horizontal heights.

\section{Summary}\label{sec:summary}
%OSC cooling rate and equilibrium beam parameters calculation in a 3D genral coupled lattice. 
%
%OSC cooling rate expressed more elegently using Courant-Snyder functions.
%
%Application of OSC for high-power EUV and soft X-ray generation.
%
%Application of OSC for ultrashort radiation pulse generation.
%
%Calculation of radiation properties.

In summary, we propose to use the longitudinal strong focusing principle to lower the beam energy spread locally in a storage ring, and use it in the Echo SSMB scheme to obtain a solution of 1 kW EUV source. 

%The potential application is expected to be wide.

%This work is supported by Tsinghua University Dushi Program. 

%We encourage readers to refer to Ref.~\cite{OSCSSMB2024} for more technical details.

%\section{ACKNOWLEDGMENTS}
%This work is supported by  the National Natural Science Foundation of China (NSFC Grant No. 12522512),  the National
%Key Research and Development Program of China (Grant No. 2022YFA1603401), the Beijing
%Outstanding Young Scientist Program (No. JWZQ20240101006) and Tsinghua University Dushi Program. 

%\begin{thebibliography}{99}   	% Use for  10-99  references

\end{document}